\documentclass[10pt]{Preamble/llcns}
\usepackage{amsmath,amssymb,amsfonts}
\usepackage{algorithmic}
\usepackage{graphicx}
\usepackage{textcomp}
\usepackage{xcolor}
\usepackage{enumitem}
\usepackage[style=base]{caption}
\usepackage{wrapfig}
\usepackage[T1]{fontenc}
\usepackage{mathtools}
\usepackage{subfigure}
\usepackage{enumitem,comment,wasysym}
\usepackage{color, soul}
\usepackage{pbox}
\usepackage{balance}
\usepackage{listings}
\usepackage{tikz,subfigure}
\usepackage{url}
\usepackage{booktabs} 
\usepackage{multirow}
\usepackage{xspace}
\usepackage{setspace}
\usepackage{hyperref}
\usepackage{hyphenat}
\usepackage{tablefootnote}
\usepackage{arydshln}
\usepackage{adjustbox}
\usepackage{tablefootnote}
\usepackage[stable]{footmisc}
\usepackage[most]{tcolorbox}
\usepackage{bibnames}
\usepackage[font=footnotesize,labelfont=bf]{caption}
\usepackage{times}
\newcommand{\name}{\textsc{PoW-How}\xspace}

\newcommand{\eg}{{e.g.,}\xspace}
\newcommand{\ie}{{\it i.e.,}\xspace}
\newcommand{\rpi}{{Raspberry Pi 3}\xspace}
\newcommand{\point}[1]{\vspace{0.01in}\par\vspace{0.01in}\noindent{\textbf{#1:} }}

\ifdefined\nocomment
    \commentfalse
\fi

\makeatletter
\renewcommand\section{\@startsection
  {section}{1}{0mm}
  {-\baselineskip} 
  {0.005\baselineskip}
{\bfseries}}
\makeatother

\makeatletter
\renewcommand\subsection{\@startsection
  {subsection}{2}{0mm}
  {-0.3\baselineskip} 
  {0.005\baselineskip}
  {\bf\normalsize}}
\makeatother

\begin{document}
\title{\name: An enduring timing side-channel to evade online malware sandboxes}

\author{Antonio Nappa$^{\ddagger\dagger\blacklozenge}$, Panagiotis Papadopoulos$^{\circ}$, Matteo~Varvello$^{\star}$,\\ Daniel Aceituno Gomez$^{\dagger}$, Juan Tapiador$^{\dagger}$, Andrea Lanzi$^{\diamond}$}
\institute{
$\ddagger$~UC Berkeley,
$\dagger$~Universidad Carlos III de Madrid,
$\blacklozenge$~Zimperium zLabs Team,\\
$\circ$~Telefonica Research,
$\star$~Nokia~Bell Labs,
$\diamond$~University of Milan
}

\maketitle
\pagestyle{plain}
\begin{abstract}
Online malware scanners are one of the best weapons in the arsenal of cybersecurity companies and researchers. A fundamental part of such systems is the sandbox that provides an instrumented and isolated environment (virtualized or emulated) for any user to upload and run unknown artifacts and identify potentially malicious behaviors. The provided API and the wealth of information in the reports produced by these services have also helped attackers test the efficacy of numerous techniques to make malware hard to detect. 

The most common technique used by malware for evading the analysis system is to monitor the execution environment, detect the presence of any debugging artifacts, and hide its malicious behavior if needed. This is usually achieved by looking for signals suggesting that the execution environment does not belong to a the native machine, such as specific memory patterns or behavioral traits of certain CPU instructions. 

In this paper, we show how an attacker can evade detection on such online services by incorporating a Proof-of-Work (PoW) algorithm into a malware sample. Specifically, we leverage the asymptotic behavior of the computational cost of PoW algorithms when they run on some classes of hardware platforms to effectively detect a non bare-metal environment of the malware sandbox analyzer. To prove the validity of this intuition, we design and implement the \name framework, a tool to automatically implement sandbox detection strategies and embed a test evasion program into an arbitrary malware sample. Our empirical evaluation shows that the proposed evasion technique is durable, hard to fingerprint, and reduces existing malware detection rate by a factor of 10.
Moreover, we show how bare-metal environments cannot scale with actual malware submissions rates for consumer services.
\end{abstract}
\section{Introduction}
\label{sec:intro}

Malware attacks have a significant financial cost, estimated around \$1.5 trillion dollars annually (or \$2.9 million dollars per minute)~\cite{malwarecost2018}, with predictions hinting at this cost to reach \$6 trillion dollars by 2021~\cite{malwareCost}. Due to the sheer amount of known malware samples~\cite{virushare,virustotalstats},  manual analysis neither scales nor allows to build any comprehensive threat intelligence around the detected cases (\eg malware clustering by specific behavior, family or infection campaign). To address this problem, security researchers have introduced \emph{sandboxes}~\cite{anubis}: isolated environments that automate the dynamic execution of malware and monitor its behavior under different scenarios. Sandboxes usually comprise a set of virtualized or emulated machines, instrumented to gather fundamental information of the malware execution, such as system calls, registry keys accessed or modified, new files created, and memory patterns.

As a next step, online services came to bring malware analysis from security experts to the common users~\cite{sandboxes}. Online malware scanners are not only useful for the users but also for the attackers. In fact by allowing an artefact to be checked multiple times against various state-of-the-art of malware analysis sandboxes, attackers can tune the evasiveness of their malware samples by exploiting the feedback reported by these services and try various techniques before making the sample capable of detecting the presence of a sandbox.
Specific CPU instructions, registry keys, memory patterns, and \emph{red pills}~\cite{1624022,fistful,kemufuzzer} are only a few of the signals used by attackers for identifying glitches of the  emulated environment that can disclose the presence of a sandbox environment. These techniques have triggered an arms-race, with the more sophisticated web malware scanners rushing to spoof any such exploitable signals~\cite{barecloud}.

In this work, we show how an attacker can evade malware analysis of these scanning services by leveraging Proof-of-Work (PoW)~\cite{powdwork} algorithms. Our intuition lies on the fact that, like NP-class problems~\cite{npp}, the asymptotic behavior of a PoW algorithm is constant in terms of computational power~\cite{powdwork}, \eg CPU and memory consumption which remain stable over time. Accordingly, PoW algorithms are perfect candidates for benchmarking the computation capability of the underlying hardware. In such scenario the benchmark can be leveraged as a fingerprint of the underlying computing infrastructure, revealing the presence of a sandbox since it shows a statistical deviation compared with the native hardware platform. Moreover, current defensive techniques that aim at spoofing the virtualization signals present in contemporary sandboxes cannot act as countermeasures against the stable timing side-channels that our technique exploits.

A key advantage of using PoW techniques is that they are a time-proof and self-contained mechanism compared to other more fine-grained timing side-channel approaches that try to detect the underlying hardware machine. In fact, our system does not require access to precise timing resources for detecting the emulated environment (\eg network or fine-grained timers). In our evaluation we empirically validate that a PoW-based technique can detect an emulated environment with high precision just by looking at the output of the algorithm (i.e., execution time, and number of successful iterations). Furthermore, 
PoW implementations do not raise any suspicion to automated malware sandboxes compared with the stalling code  (\eg infinite loops and/or sleep) that is easier to detect because of CPU idleness~\cite{10.1145/2818000.2818030}. Fingerprinting PoW algorithms as a malware component is feasible \eg by checking the usage of particular cryptographic instructions. However, using it as a proxy signal for detecting malware would produce a large number of false positives since PoW algorithms are part of legitimate applications such as Filecoin ~\cite{filecoin} and Hashcash~\cite{hashcash}.

\noindent{\bf Contributions.} 
In this paper, we make the following contributions: 
\begin{enumerate}[leftmargin=0.5cm, itemsep=0.2cm]
    \item We design and implement \name: a framework to automatically create, inject, and evaluate PoW-based evasion strategies in arbitrary programs. \name operates as a three-step pipeline. 
    First (step 1) multiple PoW algorithms are thoroughly tested across different hardware platforms (Raspberry Pi 3, Dual Intel Xeon, Intel i9), operating systems (Linux Ubuntu 18.03 and Windows 10), and machine loads. The outcome of these tests (step 2) is used to build a statistical characterization of each PoW's execution time under each setting. We use the Bienaym\'{e}–Chebyshev inequality~\cite{chebychev} to obtain statistical evidence about the expected execution time. Next, a miscreant can upload its malware to the \name framework and select the evasion mechanism to be used. Finally (step 3), \name automatically evaluates the accuracy of the evasion mechanism selected and embedded in the uploaded malware via several tests on multiple online sandbox services~\cite{sandboxes}.
    
    \item We empirically evaluate each step of \name's pipeline. For the PoW threshold estimation, we have tested three popular PoW algorithms
    (Catena~\cite{catena}, Argon2~\cite{argon2,argon2rfc} and Yescrypt~\cite{yescrypt}) using multiple configurations. During 24 hours of testing, we find Chebyshev inequality values higher than 97\% regardless of PoW and setting. This result verifies high determinism in PoW execution times on real hardware, thus validating the main intuition behind this work.  We test our technique on top of two known ransomware families by submitting to three sandboxes several variants that include PoW-based evasion. The results demonstrate how PoW-based evasion 
    reduces the number of detections, even in the presence of anti-analysis techniques such as code virtualization or packing.
    
    \item To further 
    quantify the efficacy of  PoW-based evasion with real-world sandboxes, we wrote a  fully functional malware sample, integrated with an evasion mechanism based on Argon2, and submitted it to several online sandboxes. All the reports from each sandbox mark our malware as \emph{clean}. 
    We further discuss the behavioral analysis for our malware, as well as potential countermeasures to this novel PoW-based evasion mechanism we have proposed. To ensure the reproducibility of our results and foster further research on this topic, we make the source code of our system publicly available~\cite{repo}\footnote{ \url{https://github.com/anonnymousubmission/Esorics2021_Paper159}}.
\end{enumerate}
\section{Background}
\label{sec:background}

\subsection{Malware and Malware Analysis}
Together with the evolution of malicious software, researchers and professionals have tried to improve their tools and skills to understand malware and counter its consequences. There is a huge amount of literature devoted to analyze and counter malware~\cite{CyberProbe_NDSS14,autoprobe,Gu_ACSAC09_botProber,ppipup,ppi,Wang_Oakland10_TaintScope,DBLP:conf/sp/KolbitschHKK10,DBLP:conf/sp/MoserKK07}. Every aspect of this phenomenon has been taken into consideration, from its network infrastructure, to the code that gets reused among samples, unexplored paths in the control-flow, sandbox design and instrumentation. Nonetheless the arms race keeps running, while new analysis evasion techniques are found, new countermeasures get developed. 

\point{Anti-Analysis Techniques}
There are several anti-analysis techniques which have been developed during the years by miscreants, and promptly countered by our community: \eg packers~\cite{omniunpack,ugarte}, emulators~\cite{rotalume}, anti-debugging and anti-disassembly tricks and stalling code. Among all these techniques the only one that seems to resist is stalling code, which is very difficult to detect~\cite{lastlinestall}. Indeed, over 70\% of all malware attacks involved evasive zero-day malware in Q2 of 2020: a 12\% rise on the previous quarter~\cite{malware2020}. This denotes that evasive malware is a phenomenon that will hardly disappear and there will always be  continuous research in evading analysis systems. 

\subsection{PoW for Malware Analysis Evasion}
Proof-of-Work (PoW)~\cite{powdwork} is a consensus mechanism that imposes computation workload on a node. A key feature of such algorithms is their asymmetry: the work imposed on the node is moderately hard but it is easy for a server to check the computed result. There are two types of PoW protocols: (a) \emph{challenge-response} protocols, which require an 
interactive link between the server and the client, and (b) \emph{solution-verification} protocols, which allow the client to solve a self-imposed problem and send the solution to the server to verify the validity of the problem and its solution. Such 
PoW protocols (also known as CPU cost functions) leverage algorithms like hashcash with doubly iterated SHA256~\cite{laurie2004proof}, momentum birthday collision~\cite{larimer2014momentum}, cuckoo cycle~\cite{tromp2015cuckoo}, and more.

In \name we use Argon2, which guarantees that by using the same input parameters, the amount of computation performed is asymptotically constant; hence, the variance of Argons2's execution time $T$ is very small on the same platform. Moreover, Argon2 is based on a memory-hard function which, even in the case of parallel or specialized execution (\eg ASICs or FPGAs), will not enhance scalability, and hence remains computationally bounded due to its asymptotic behavior. 

The Argon2 algorithm takes the following input:
 \begin{itemize}
     \item A message string $P$, which is a password for password hashing applications. Its length must be within 32-bit size.
     \item A nonce $S$, which is used as salt for password hashing applications. Its length must be within 32-bit size. 
     \item A degree of parallelism $p$ that determines how many independent (but synchronized) threads can be run. Its value should be within 24-bit size (minimum is 1).
     \item A tag, which length should be within 2 and 32-bit. 
     \item A memory size $m$, which is a number expressed in Kibibytes.
     \item A number of internal iterations $t$, which is used to tune the running time independently of the memory size. Its value should be within 32-bit size (minimum is 1).
 \end{itemize}

 These input parameters are used in our framework to define the computational boundary of the algorithm execution on a specific class of hardware machines. Once the parameters are set, the output of the PoW algorithm only depends on the hardware platform. 

\subsection{Side-channel Measurement}
Various techniques have been proposed to detect if applications are running inside a sandbox/virtualizer/emulator. The most reliable of them is based on timing measurements~\cite{timingfoundational}. Indeed, fine grained timers help also to build micro-architectural attacks such as Spectre and Meltdown~\cite{meltdown,spectre}. 
The intuition behind our work is that PoW algorithms offer strong cryptographic properties with a very stable complexity growth, which make the approach very resilient to any countermeasure, such as using more powerful bare-metal machines to enhance performance and reduce the space for time measurements.

By exploiting the asymptotic behavior of the PoW algorithms, we build a statistical model that can be used to guess the class of environment where the algorithm is running and consequently distinguish between physical and virtualized, emulated or simulated architectures, like different flavors of malware sandboxes.
Indeed, even fine grained red-pills techniques~\cite{fistful} such as CPU instruction misbehavior can be easily fixed in the sandbox or spoofed to thwart evasion techniques. On the other hand PoW stands on top of well defined mathematical and well defined computational behavior. Moreover, a simple modification of the PoW library avoids the malware sample to be fingerprinted by static techniques. If we take as an example of PoW complexity the one that is run in the crypto currency environment, we know that by design the computation complexity of the algorithm is increased for each new block of the blockchain transaction~\cite{bitcoin}. Such an increase of computation shows the asymptotic behavior that can be exploited by our technique. By applying PoW as a malware sandbox evasion technique, we get an off-the-shelf technique which improves the malware resilience and limits its analysis. 
\section{Our Approach: \name}
\label{sec:framework}

This section describes our threat model before describing our approach in detail. We first provide an overview of the technique (Section~\ref{sec:framework:overview}) and its main workflow. We then describe how the key parameters are estimated (Sections~\ref{sec:framework:measurem} and \ref{sec:framework:estim}) and how an arbitrary sample can be equipped with the evasion module (Section~\ref{sec:framework:integration}).

\subsection{Threat Model}
\label{sec:threat}
In this paper, we assume a malware scanning service based on virtualized or emulated sandboxes, which allows users to upload and scan their individual files for free as many times as they need. Such a service joins together results from various state-of-the-art malware analysis sandboxes before responding back to the user with a detailed report about the detection outcome of each and every sandbox scanner used.  

On the other hand, we assume an attacker who developed a program that includes (i) some malicious payload along with (ii) a technique to pause or alter the execution of the malicious program itself, when a possible malware analysis environment is detected. Before distributing the malicious program to the victims, the attacker may use a malware scanning service to assess its evasiveness.

\subsection{System Design}
\label{sec:framework:overview}
\begin{figure}[t]
\centering
\includegraphics[width=\columnwidth]{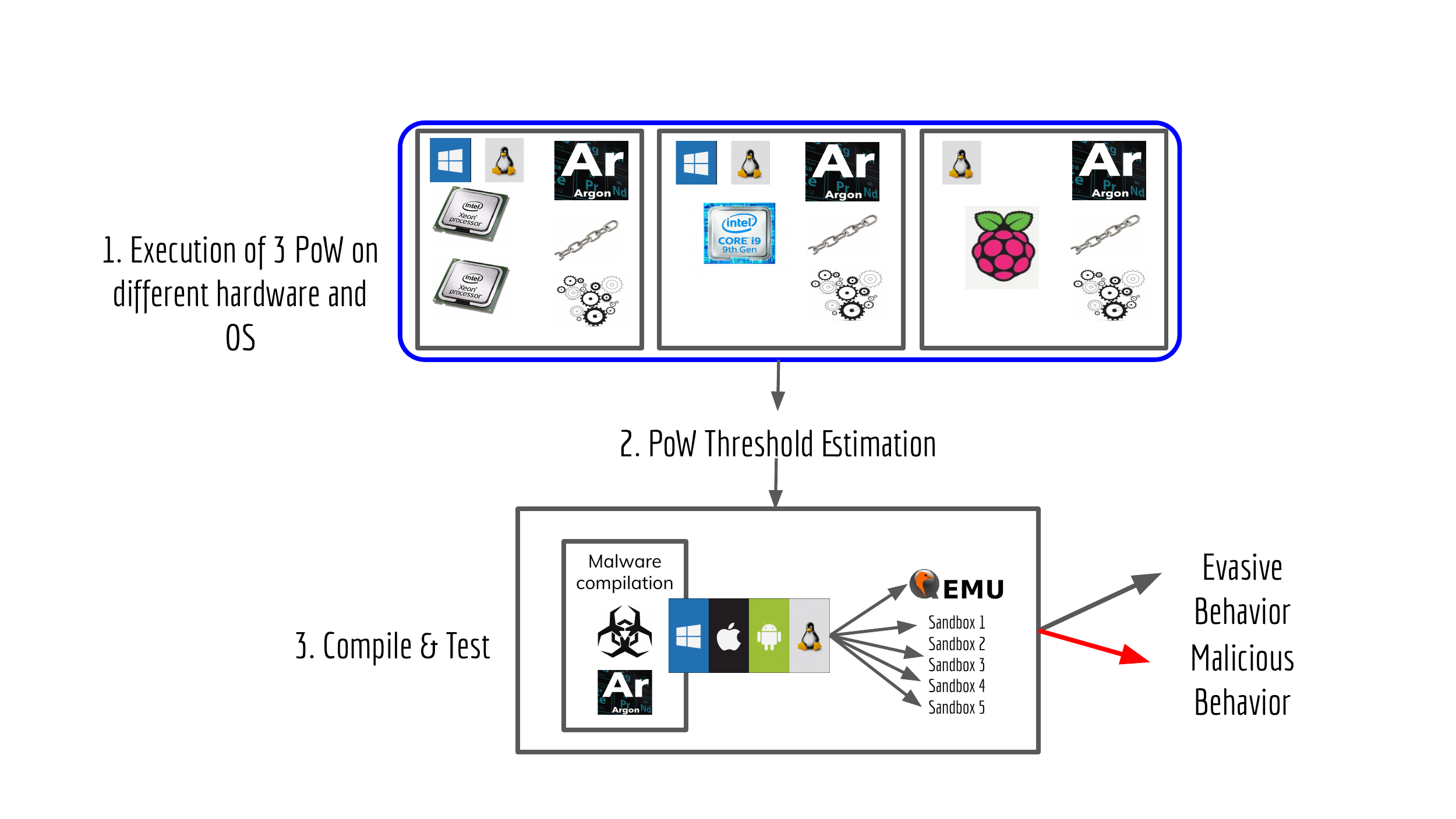}
    \caption{High level overview of \name. Step 1: execution of the PoW 
    on several hardware/OSes using different configuration settings and system load. Step 2: threshold estimation based on execution time per configuration/architecture. Step 3: malware integration and test.}
\label{fig:arch}\vspace{-0.4cm}
\end{figure}

As described in Section~\ref{sec:background}, PoW puzzles have moderately high solving cost and a very small verification time, like problems in the NP complexity class~\cite{npp}. This implies that their asymptotic behavior is constant in terms of computational cost~\cite{powdwork}, \eg CPU and memory consumption. \name exploits this asymptotic behavior to build a statistical model that can be used to identify the class of hardware machines where the algorithm is running. Such a model can later be used to distinguish between physical and virtualized architectures, like those present in  malware sandboxes. \name is a three-step pipeline (see Figure~\ref{fig:arch}):
\begin{enumerate}
    \item \textit{Performance Profiling}. It 
    executes multiple PoW algorithms on several hardware 
    and operating systems using different configuration settings and system loads.
    \item \textit{Model estimation}. The previous step provides the system with a measurement of the amount of time needed to execute the PoW on real hardware. By using the Bienaym\'{e}–Chebyshev~\cite{chebychev} inequality, it then estimates the time (threshold) expected for a particular configuration to run on a given architecture.
    \item \textit{Integration}. Once the models are built, a malware developer can select a specific PoW and parameters to associate with an arbitrary malware sample. \name then generates a module with the chosen PoW, which is integrated with the sample by building a single statically-linked executable.
\end{enumerate}
As ground truth, our methodology leverages a custom Cuckoo Sandbox~\cite{cuckoo} and  popular crowd-sourced malware scanning services (like VirusTotal or similar~\cite{sandboxes}), as a testbed to report on the accuracy of the evasiveness of the malware in real-world settings.

\subsection{Performance Profiling}
\label{sec:framework:measurem}
The first step in \name's pipeline produces a number of PoW executions using different algorithms, parameters, hardware, operating systems, and load settings:

\point{Hardware}
\name leverages three machines representative of low, medium, and high-end platforms. The high-end machine is a desktop equipped with an Intel(R) Core(TM) i9-9900X CPU @ 3.50GHz with 10 physical cores and 20 threads equipped with a PCI-e M2 512GB disk and 32 GB of RAM. The medium-end machine is a workstation equipped with a Dual Intel(R) Xeon(R) CPU E5-2643 0 @ 3.30GHz with 16 physical cores and 64GB of RAM. Finally, the low-end device is a \rpi which comes with a quad core ARMv7 Processor rev 4 (v7l) and 1GB of RAM.

\point{Systems and loads}
With the exception of the \rpi, the other hardware platforms are setup in dual boot, supporting both Linux Ubuntu 18.04.3 (64 bits) and Windows 10 (64 bits). Each platform can be further configured in \emph{idle} and \emph{busy} mode. The latter is achieved using \texttt{iperf}~\cite{perf} a CPU bound network traffic generator to keep the operating system and the CPU occupied. 

\point{PoW and parameters}
\name currently supports three popular PoW algorithms: Catena~\cite{catena}, Argon2~\cite{argon2,argon2rfc}, and Yescrypt~\cite{yescrypt}.
Each PoW algorithm is executed multiple times with different input parameters on each hardware platform, operating system, and load setting. The parameters of each algorithm allow to control the amount of memory, parallelism, and complexity of the PoW. Our selection is based on common configuration of COTS hardware devices, with respect to memory and CPU. However, not all the selected algorithms have these parameters available for tuning and in some cases, their tuning is more coarse grained~\cite{catena}.

\subsection{Threshold Estimation}
\label{sec:framework:estim}
The second step in \name's pipeline aims at estimating the PoW thresholds for different settings (PoW algorithm, parameters, hardware, operating system, and load). This is achieved through a statistical characterization of the execution time in each setting using the Bienaym\'{e}–Chebyshev inequality~\cite{chebychev}. This is a well-known result in probability theory stating that for a large class of distributions, no more than $\frac{1}{k^2}$ values of a distribution $X$ can be more than $k$ standard deviations ($\sigma$) away from the mean ($\mu$):

\begin{equation}
Pr(|X - \mu| \geq k\sigma) \leq \frac{1}{k^2}
\end{equation}

Using the empirical distribution of execution time observed in the previous step, this inequality allows us to select a threshold $T$ (\ie a maximum execution time) which guarantees a high sample population coverage. The previous deduction enables us to determine with high probability the time $T$ it will take for a PoW to run if the underlying platform is not virtualized. To reduce false positives, the evasion rule can be generalized to ``the execution environment is virtualized if the PoW does not complete $N$ executions in less than $T$ seconds.''

\subsection{Malware Integration and Testing}
\label{sec:framework:integration}
The final step in \name's pipeline is PoW integration with a malware sample provided as input. At this step, the attacker can upload its sample to \name and select the PoW-based evasion mechanism to be used,  along with its parameters. \name further informs the attacker about the predicted accuracy of this selection.

\name integrates the uploaded malware with the PoW selected and the Boost C++ libraries~\cite{boost}, which ease the OS interaction to build a single statically-linked executable. The compilation stage is automated as an Ansible~\cite{ansible} playbook and clang~\cite{clang}. The integration is achieved at linking stage, so the malware will have a stub call to an external symbol that will be linked with the chosen PoW. \name's pipeline then starts the Ansible scripts, which runs some tests and launch the compilation of the final binary for multiple platforms automatically.

\point{Testing}
To evaluate the accuracy of the newly generated evasion mechanism, we rely both on a local sandbox---a custom Cuckoo Sandbox~\cite{cuckoo} equipped with Windows 10 (64 bits), which is the most targeted OS for malware campaigns~\cite{wannacry}---and several on-line free-of-charge sandbox services~\cite{sandboxes}. Once this step is completed, \name offers to the user access to the set of reports generated by each sandbox.

\section{Evaluation}
\label{sec:eval}

In this section, we evaluate \name's pipeline. We first analyze the combination of PoWs and their parameters currently supported by \name. The outcome of this evaluation are the parameters $N$ (cycle of execution made in less than $T$ second) and $T$ (maximum execution time) to be associated with the malware sample. We then discuss the accuracy of our evasion mechanism across using various case studies across three public malware scanning services: (~\cite{joesandbox}, ~\cite{hanalysis},~\cite{virustotal}), along with our own Cuckoo Sandbox instance. 

\begin{table}[t]
\centering
\makebox[0pt][c]{\parbox{\textwidth}{%
    \begin{minipage}[t]{0.48\textwidth}
\centering
\scriptsize
    \begin{tabular}{lrrrr}
    \toprule 
        \bf Platform  & \bf Status &  \bf Win 10 & \bf Ubuntu 18.03  \\
        \midrule
        Intel i9 & \emph{idle}   & 4,500 & 9,325 \\
        & \emph{busy}            & 3,642 & 8,867\\ 
        \midrule
        Dual Intel Xeon & \emph{idle} & 6,005 & 7,897\\
        & \emph{busy}            &    4,320 & 7,012\\ 
        \midrule
        \rpi   & \emph{idle}     &     -  & 300\\
        & \emph{busy}            &     -  & 143 \\
    \bottomrule
    \end{tabular}
    \caption{Number of consecutive PoW executions per hardware and OS combination over 24 hours. For a given platform, the first line refers to results obtained with the \emph{idle} setting, while the second line refers to \emph{busy} setting.}
    \label{tab:hw_totals1}
\end{minipage}
\hfill
\begin{minipage}[t]{0.48\textwidth}
    \centering
    \scriptsize
    \begin{tabular}{lrrrrrr}
    \toprule
    \bf Garlic Graph Size   & \bf Min & \bf Max & \bf Sigma & \bf Mean & \bf K &\bf Chebyshev \\
     \midrule
    15 & 0.12 & 5.35 &   0.503 &  0.209 & 9.99 &          99.00\% \\
    18 & 1.13 & 35.61 &   4.22 &  1.86 & 7.94 &          98.41\% \\
    20 & 5.11 & 165.57 &  19.01 &  8.26 & 8.26 &          98.54\% \\
    \bottomrule
    \end{tabular}
    \caption{Statistical measurement results for Catena.}
    \label{tab:catena}
\end{minipage}
}}\vspace{-0.5cm}
\end{table}

\subsection{Threshold Estimation and PoW Algorithm Choice}
\label{lab:estimation}
For each PoW, we have selected different configurations with respect to memory footprint, parallelism, and algorithm internal iterations (see Tables~\ref{tab:catena} for Catena and ~\ref{tab:argon-yescrypt} for Argon2i and Yescryot). Argon2i and Yescrypt have similar parameters (memory, number of threads, blocks) whereas Catena's only parameter is a graph size which grows in memory and will make its computation harder as the graph size increases.

\name executes each PoW configuration on the low-end (\rpi), medium-end (Dual Intel Xeon), and high-end (Intel i9) machines. All PoW configurations are executed sequentially during 24 hours on each machine for both idle and busy conditions. As pointed out in Section~\ref{sec:framework}, with the exception of the \rpi, all tests are performed on two operating system per hardware platform: Linux Ubuntu 18.04.3 (64 bits) and Windows 10 (64 bits).

Table~\ref{tab:hw_totals1} shows the total number of PoW executed over 24 hours per hardware, operating systems, and CPU load (\emph{idle} or \emph{busy}). Regardless of the CPU load on each machine, we observe two key insights. First, there is a significant drop in the number of PoW executions when considering Linux vs Windows, which is close to a 50\% reduction in the high-end machine. This is due 
to operating system interaction, ABI and binary format, and ultimately idle cycle management. Second, a 30x reduction in the number of PoW executions when comparing high-end and low-end platforms, \eg under no additional load the \rpi completes 300 executions versus an average of 8,611 executions on both the high and medium-end machines. Finally, extra load on the medium and high-end machines  causes a reduction in number of proofs computation of about 6-10\%, averaging out to 7,300 executions between the two machines. A more dramatic 50\% reduction was instead measured for the \rpi.

Next, we statistically investigate PoW execution times by mean of the Bienaym\'{e}–Chebyshev inequality (see Section~\ref{sec:framework:estim}). To balance equally sized datasets, we sampled 150 random executions (\ie the total number of executions that were possible to complete on the low-end platform) from the 9,325 executions available from both the medium and high-end platforms.  
Tables~\ref{tab:catena} and~\ref{tab:argon-yescrypt} show for each PoW and configuration, several statistics (min, max, $\sigma$, and $K$, Chebyshev inequality) of the PoW execution time computed across hardware platforms, OSes (when available), and load condition (idle, busy). Overall, we measured Chebyshev inequality values higher than 97\% regardless of the PoW and its configuration. This confirms high determinism in the PoW execution times on real hardware, validating the main intuition behind this work. 

\point{Algorithm choice}
The results above provide the basis to select a PoW algorithm along with its parameters to integrate with the input malware sample. These results indicate  that PoW selection has minimal impact on the expected accuracy of the proposed evasion mechanism. We then selected Argon2i (with 8 threads, 100 internal functions and 4KiB of memory) motivated by its robustness and maturity.
We leverage the results from Table~\ref{tab:argon-yescrypt} (top, second line) to set the parameters $N$ (PoW execution) and $T$ (evasion threshold) of an  Argon-based evasion mechanism. The table shows that $K=8.1$ seconds allows a good coverage for the execution time population (98.3\%). We  opted for a more conservative value of $T=10$ and further performed  multiple tests on our internal Cuckoo Sandbox. Given that our Cuckoo Sandbox could not even execute 1 PoW with $T=10$, we simply set $N>1$. We will use this configuration for the experimentation described in the remaining of this paper.

\subsection{Case Study: Known Malware}
We first analyze the effect of adding our PoW-based evasion strategy to the code of two well-known ransomware samples: Relec and Forbidden Tear.  The use of real-world malwares, which are well know and thus easy to detect, allows us to comment on the impact that PoW-based evasion has on \textit{malware reuse}, the practice of recycling old malware for new attacks. We use \name to generate various combinations of each original ransomware with/without PoW-based evasion strategy, code virtualization\footnote{This cannot be applied to ForbiddenTear since it is written in .NET.}, and packing offered by Themida, a well-known commercial packer~\cite{themida}. We verify that all the malicious operations of the original malwares were preserved across the generated versions. 

\begin{table}[t]
\centering
\makebox[0pt][c]{\parbox{\textwidth}{%
    \begin{minipage}[t]{0.48\textwidth}    \centering
    \scriptsize
    \begin{tabular}{lrrrrrrrr}
    \toprule
     \bf Thr. & \bf It. & \bf Mem. & \bf Min & \bf Max & \bf Sigma & \bf Mean & \bf K & \bf Cheb.\\
    \midrule
     1 & 10   & 1KB  & 0.01 & 0.70 &  0.09 &  0.02 & 7.9 & 98.4\% \\
     8 & 100  & 4KB  & 0.20 & 9.28 &  1.07 & 0.46 & 8.1 & 98.3\% \\ 
     16 & 500 & 8KB  &  2.03 & 88.8 &  10.5 &  3.85 & 7.9 & 98.4\% \\
    \midrule
    \midrule
    1 & 1K    & 8KB  & 0.00 & 0.02 &   0.00 &  0.01 & 6.1 &  97.3\% \\
    8 & 2K    & 32KB & 0.03 & 0.56 &   0.05 &  0.05 & 10.5 & 99.1\% \\
    16 & 4K   & 64KB & 0.08 & 5.00 &   0.51 &  0.19 & 9.4 & 98.9\% \\
    \bottomrule
    \end{tabular}
    \caption{Statistical measurement results for Argon2i (top) and Yescrypt (bottom). Thr. = number of threads. It. = number of algorithm steps. Mem. = amount of memory used in KiB. Cheb. = Chebyshev coverage.}
    \label{tab:argon-yescrypt}
\end{minipage}
\hfill
\begin{minipage}[t]{0.48\textwidth}    \centering
    \centering
    \scriptsize
    \begin{tabular}{lrrr}
    \toprule
      \bf Test & \bf Relec & \bf Forbidden  & \bf Hello \\
      & & \bf Tear & \bf World \\
        \midrule
        Original                    & 23/72   & 26/72  & 3/72\\ 
        \begin{tabular}{@{}l@{}}Original+Code Virtualizer\end{tabular}& 32/72   & n/a   & 19/72 \\
        Original+Themida            & 33/72   & 21/72  & 17/72 \\
        \begin{tabular}{@{}l@{}}Original+PoW+Code Virtualizer\end{tabular} & 29/72   & n/a   & 0/72 \\ 
        Original+PoW+Themida        & 32/72   & 18/72  & 9/72 \\ 
        \bf Original+PoW                & \bf 3/71   & \bf 3/72  & \bf 2/72  \\ 
    \bottomrule
    \end{tabular}
    \caption{Online Sandbox detection results for 2 ransomware samples (Relec and Forbidden Tear) and a benign test program using various anti-analysis configurations.}
    \label{tab:caseStudies}
    \end{minipage}
    }}\vspace{-0.5cm}
\end{table}

We submitted all malware variants to three online sandboxes for analysis and checked how many AV engines (antivirus products) flag each variant as malicious (see Table~\ref{tab:caseStudies}). In the case of Relec, adding code virtualization or packing, results in more AV engines detecting the sample as malicious. This is likely due to the engines flagging such protections, not the malware sample itself. In all cases, the addition of PoW decreases the number of detections by a factor of 10~\cite{vtrelecnostrings}, reaching a level where the difference between the label \textit{malicious} and \textit{false positive} is evanescent. 

Table~\ref{tab:caseStudies} also show results when submitting several variants of a standard Hello World program. Note that the original code has been flagged as malicious by 3 AV engines, though as it is possible to see from the report the detections are mislabeled \ie{ Relec is not recognized}. This false positive could be due to a large number of submissions of the same code hash (due to its simplicity and popularity), our source IP being flagged, and other unknown factors which may influence the scoring. The table also shows that adding code virtualization or packing translates into a substantial increase in false positive detections even of a simple Hello World program, confirming our intuition above. Instead, adding our PoW-based evasion strategy results in less false positives, one less than the original code. This is likely due to the fact that our code on top of Hello World has more entropy, respect to a very simple one line program, looking more legit to engines that measure such kind of parameters.

Overall, these three case studies show that a PoW-based evasion strategy reduces the number of detections by 10x with known malware by preventing the sample from executing in the analysis sandbox. This result demonstrates large potential for malware reuse by coupling it with PoW-based evasion strategy. In the next section, we perform more controlled experiments based on \textit{fresh} (\ie previously unseen) malware.

\begin{figure}[t]
    \centering
    \begin{minipage}[t]{0.32\textwidth}
        \centering
        \includegraphics[width=\columnwidth]{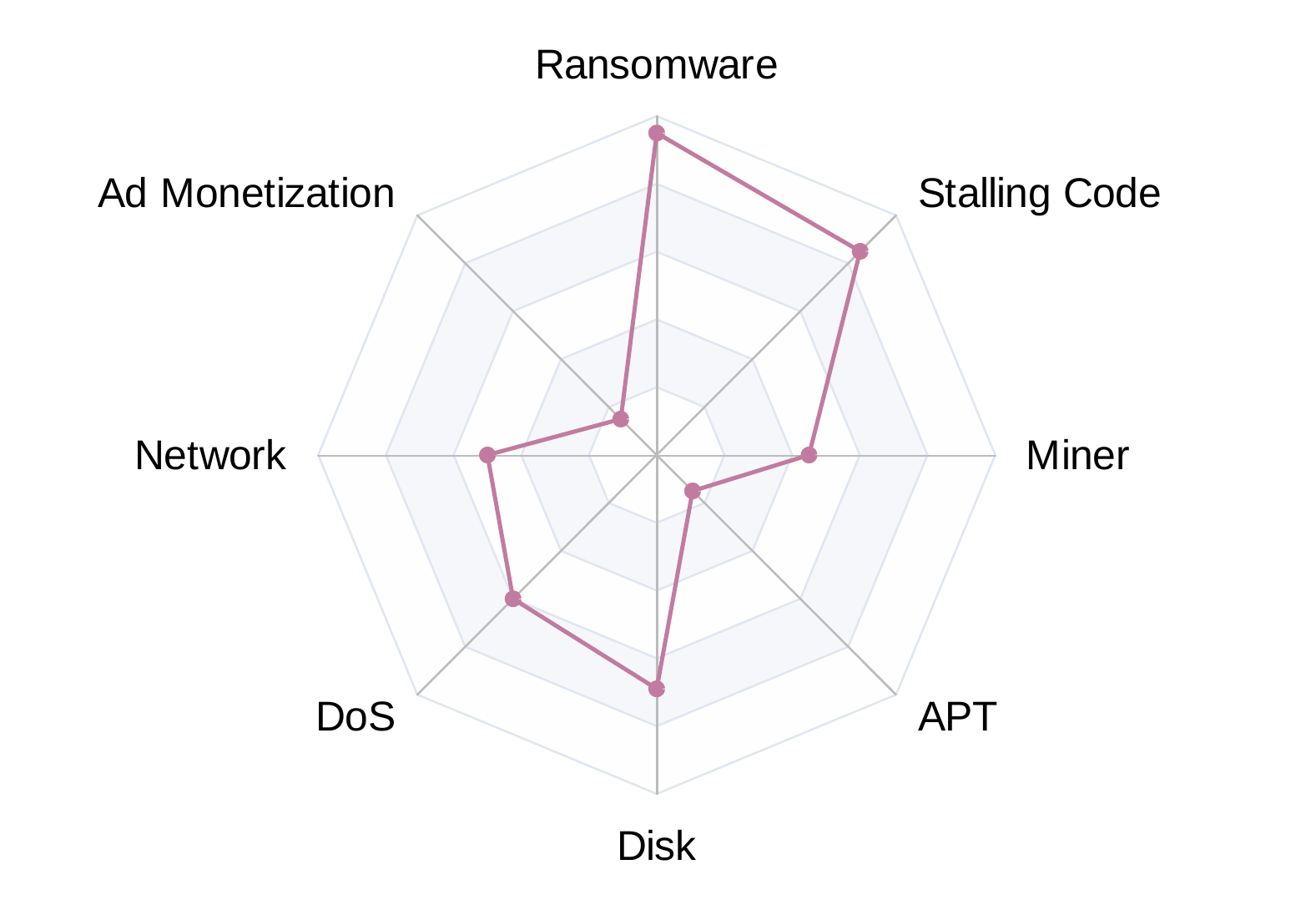}
        \caption{Behavioral map of the malware PoC \emph{without} PoW and \emph{without} full static protection enabled.}
        \label{fig:nostrip}
    \end{minipage}
    \hfill
    \begin{minipage}[t]{0.32\textwidth}
        \centering
        \includegraphics[width=\columnwidth]{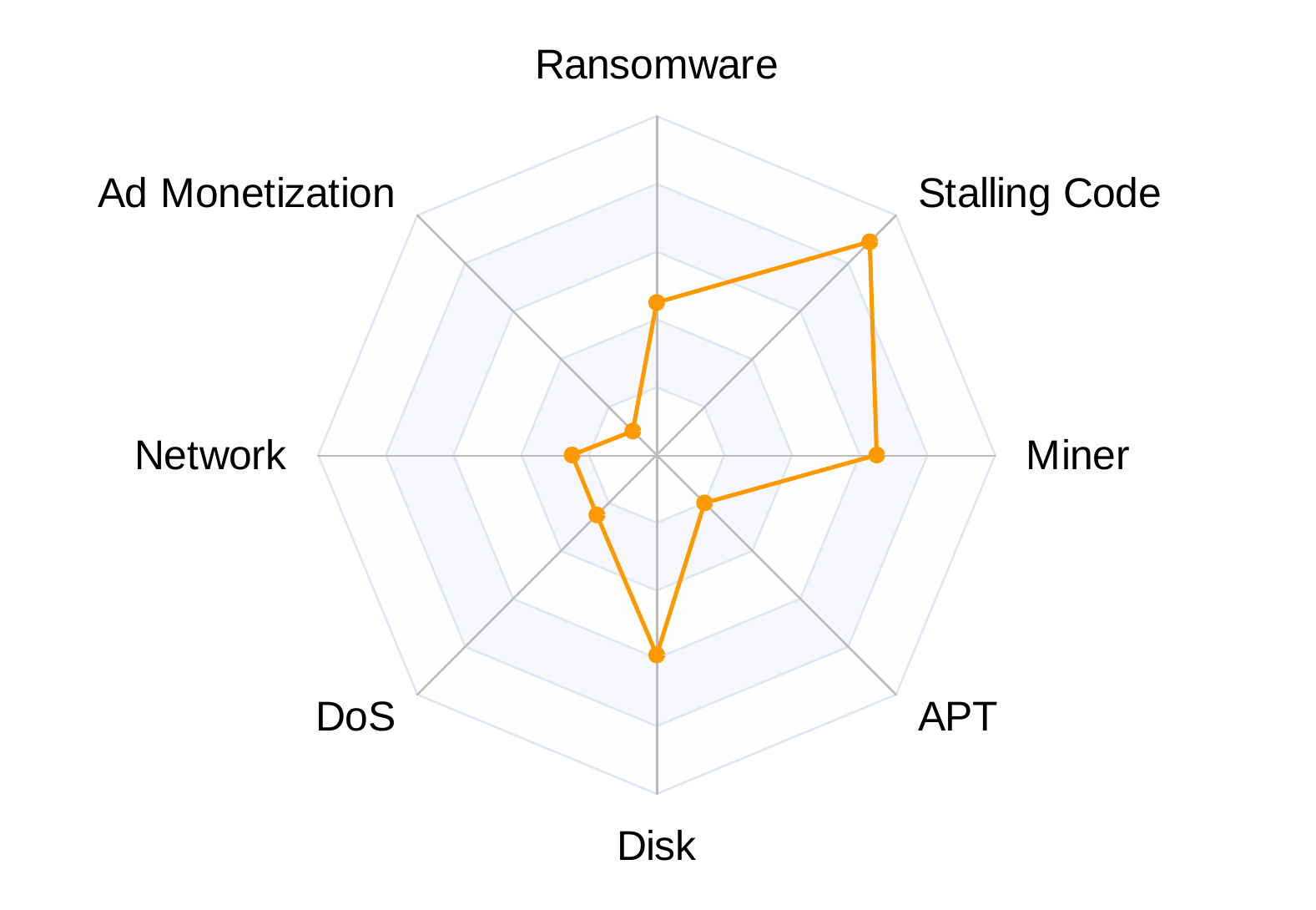}
        \caption{Behavioral map of the malware PoC  \emph{without} PoW and \emph{with} full static protection enabled.}
        \label{fig:evader}
    \end{minipage}
    \hfill
    \begin{minipage}[t]{0.32\textwidth}
        \centering
        \includegraphics[width=\columnwidth]{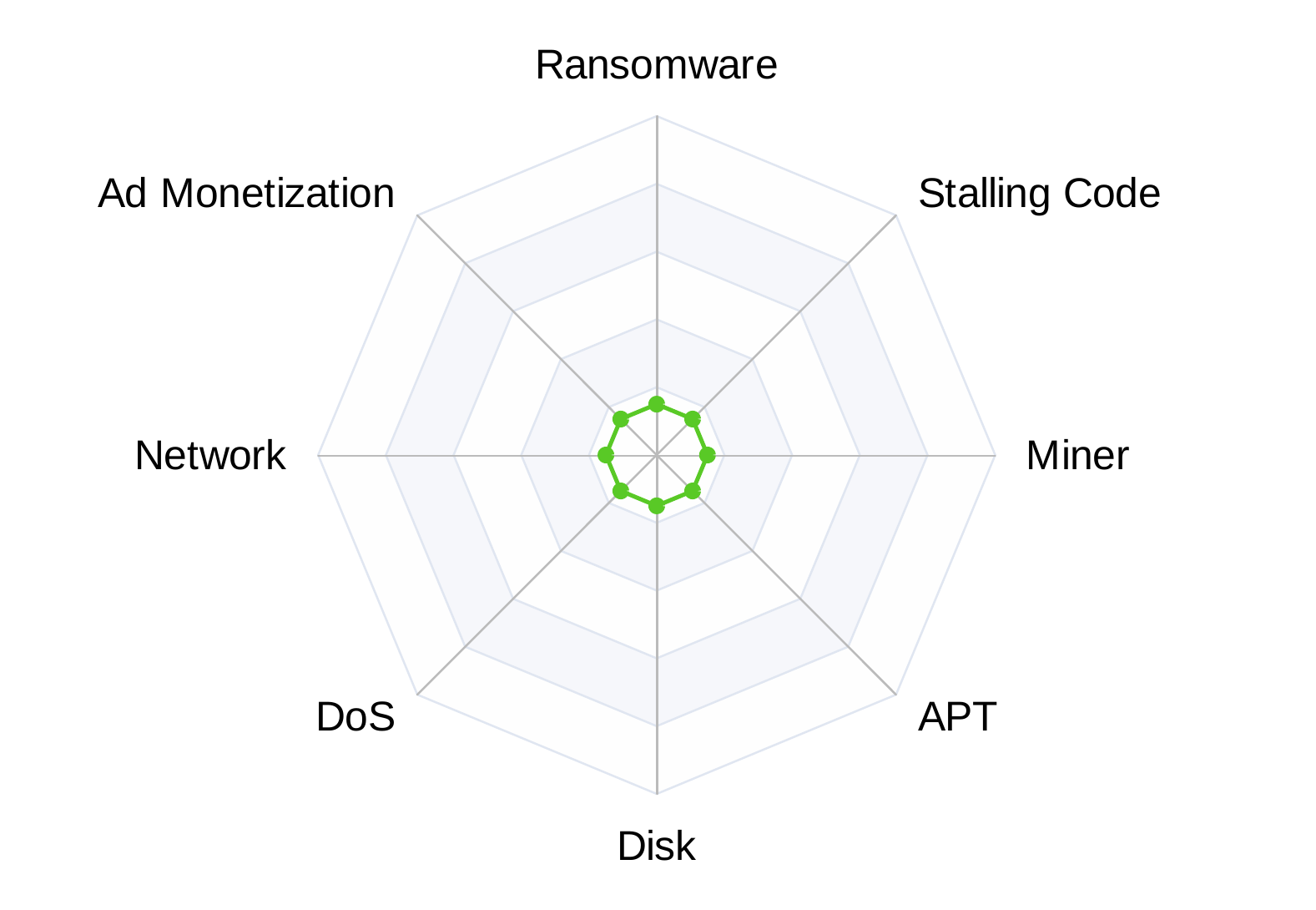}
        \caption{Behavioral map of the malware PoC  \emph{with} PoW and \emph{with} full static protection enabled.}
        \label{fig:naked}
    \end{minipage}\vspace{-0.4cm}
\end{figure}

\subsection{Case Study: Fresh Malware Sample}
In order to further explore the results obtained in the previous case studies, we wrote a simple malware PoC (roughly~150 LoC) for Windows 10 (VC++) and Linux (C++). Our malware sample implements a basic ransomware functionality which scans the entire hard drive and encrypts all its files. This behavior should be easy to detect by any malware scanning services.\footnote{The malware detection report for this malware without our PoW-based evasive measure has been anonymized~\cite{noteva,mal_vt}.} Using \name, we automatically embed a PoW (Argon2i, as we will discuss below) and make sure to exhibit its malicious activity only if the PoW is successfully executed at least $N$ times before a timeout $T$. 
Finally, we submitted different variants of our malware sample (with PoW, without PoW, with static sanitization) 
to several on-line sandboxes and the results were disheartening (see Table~\ref{tab:malsand}). For the static sanitation we remove the symbol tables and debugging symbols. Note that very similar results were also achieved with our local Cuckoo Sandbox.
It is important to note that to check the execution of the malware payload
we insert a \emph{create-file} function at the beginning of the malware payload
itself. Such file creation is visible on the behavioral report of the
analyzed sandboxes in case the malware payload is executed\footnote{This reference has been anonymized not to violate the terms of service of sandbox vendors~\cite{noteva}}. We used such a simple test to check whether the PoW algorithm detects
the emulated environment and so validate our technique. In case such a file is not present on the
behavioral report, it means the PoW algorithm detects the emulated environment and
stops the payload execution.
None of the analyzed sandboxes is able to execute more than 1 PoW during $T=10$ (or even $T=20$ sec), which is worse than what a \rpi can do even in presence of some extra load (\eg see max value in the top of Table~\ref{tab:hw_totals1}).

We made all the reports of our analysis publicly available, including screenshots of evasive malware samples~\footnote{The references have been anonymized not to violate the terms of service of sandbox vendors~\cite{eva1,eva2,eva3,eva4,eva5,evavt,evavtsgamo}}. It has to be noted that not all sandboxes report are the same, but they all signal the hard drive scan (Ransomware behavior) without full static protection (i.e., with the default compiler options). In Table~\ref{tab:malsand} the number of PoW executed is visible only if a screenshot of the sandbox is available. As for the sandbox execution timeout, not all the analysis services had it available for selection.

\point{Detection Rate Decrease}
As it is possible to see \name's approach is capable of reducing to zero the detection rate of roughly 70 antiviruses run by the tested sandboxes~\cite{virustotal,joesandbox,hanalysis} for any sample that we have tested. We have investigated the multiple facets of our technique (static and dynamic). Thus we conclude after looking also at the behavioural results of our samples that the whole technique is capable of reducing the detection rate to zero. The behavioural part plays a fundamental role as it is possible to see from the Hello World example and the behavioural maps generated by AV labels of Figures~\ref{fig:nostrip}-\ref{fig:naked}.



\section{Security Analysis}
\label{sec:analysis}

The results shown in the previous section demonstrate that a \name-ed malware can effectively detect a sandbox and abort the execution of any malicious payload. This strategy is effective in getting a malware sample marked as ``clean'' by all sandboxes tested by \name (see Table~\ref{tab:malsand}). \textbf{\name's technique is simple to deploy, it does not require precise timing measurements and, thanks to its algorithmic properties, it will last for many years as a potential threat.}

We next discuss in detail the \emph{behavioral} analysis of our malware. This is an analysis produced by a sandbox related to how a malware interacts with file system, network, and memory. If any of the monitored operations matches a known pattern, the sandbox can raise an alarm.  

\begin{table}[t]
\centering
\scriptsize
    \begin{tabular}{lrrrrr}
    \toprule
    \bf Sandbox & \bf Evasion Timeout & \bf PoW  Timeout &\bf \# of PoW executed & \bf Timeout  & \bf Notes \\
    \midrule
    Sandbox1 & $10$ secs & 50 & 1 & 120 & Clean\\
    Sandbox1 & $15$ secs & 45 & 1 & 180 & Clean\\
    Sandbox1 & $20$ secs & 40 & 1 & 240 & Clean\\
    Sandbox1 & $20$ secs & 15 & 1 & 500 & Clean\\
    Sandbox2 & $20$ secs & 15 & 0 & N/A & Clean\\
    Sandbox3 & $20$ secs & 45 & N/A & N/A & Clean\\
    Sandbox3 & $20$ secs & 15 & N/A & N/A & Clean\\
    \bottomrule
    \end{tabular}
    \caption{Execution results of a custom ransomware sample on various sandboxes}
    \label{tab:malsand}\vspace{-0.6cm}
\end{table}

Figures~\ref{fig:nostrip},~\ref{fig:evader}, and~\ref{fig:naked} show the behavioral analysis of our malware on a radar plot, labelled with most prevalent AV labels. The samples were submitted with different combinations of PoW and static protection. In Figure~\ref{fig:nostrip}, the radar plot is mostly ``green'' (benign) with respect to some operations like phishing, banker and adware for which we would not expect otherwise. However, four ``suspicious'' (orange) behaviors are reported with respect to evader, spyware, ransomware, Trojan operations. While our malware PoC is not labeled as ``malicious'' (red), the suspicious flags for our binary would trigger further manual analysis that coukd reveal its  maliciousness. It is thus paramount to investigate and mitigate such suspicious flags. 

Our intuition is that the suspicious flags are due to the fact that our malware is neither packed nor stripped, and hence some of its functionality \ie exported functions, linked libraries, and function names are visible through basic static analysis that is usually also implemented in the dynamic sandbox environment. 
Accordingly, we strip out the  whole static information from our binary and resubmit it as a new binary. Figure~\ref{fig:evader} shows the behavioral analysis of our PoC malware without PoW-based sandbox detection but with full static protection enabled. As expected, various signals have dropped from the behavioral report. Finally, Figure~\ref{fig:naked} shows the result of adding PoW to the last binary. A completely green radar plot which does not raise any suspicion illustrates the evasion effect of \name. 

\begin{figure}[t]
\centering
\begin{minipage}[t]{0.32\textwidth}
\centering
    \includegraphics[width=1\columnwidth]{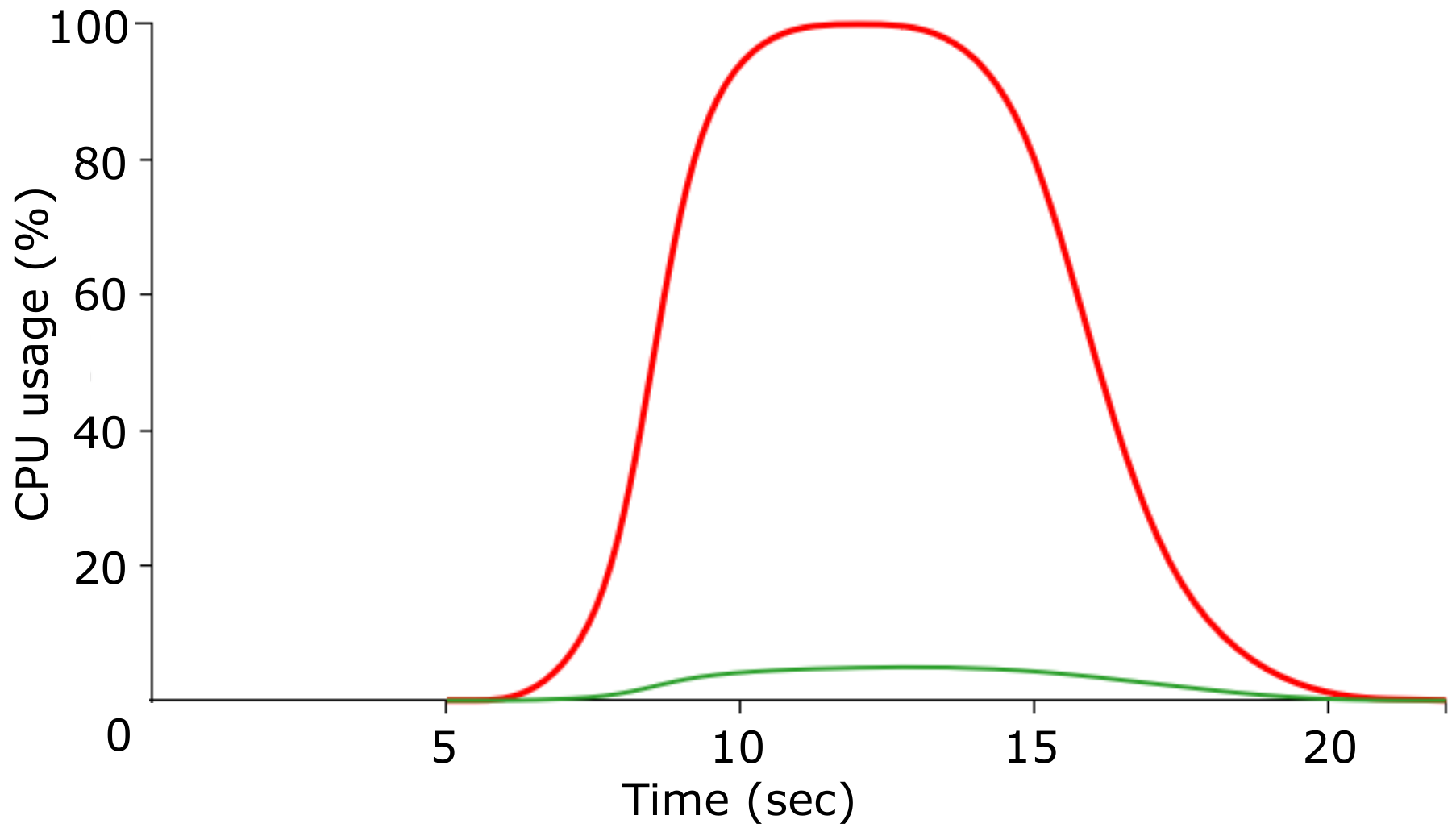}
    \caption{CPU consumption  of our malware PoC (Argon2d) Malware:red line, System Idle (PID 0):green line.}
    \label{fig:cpueva}
\end{minipage}
\hfill
\begin{minipage}[t]{0.32\textwidth}
    \centering
    \includegraphics[width=1\columnwidth]{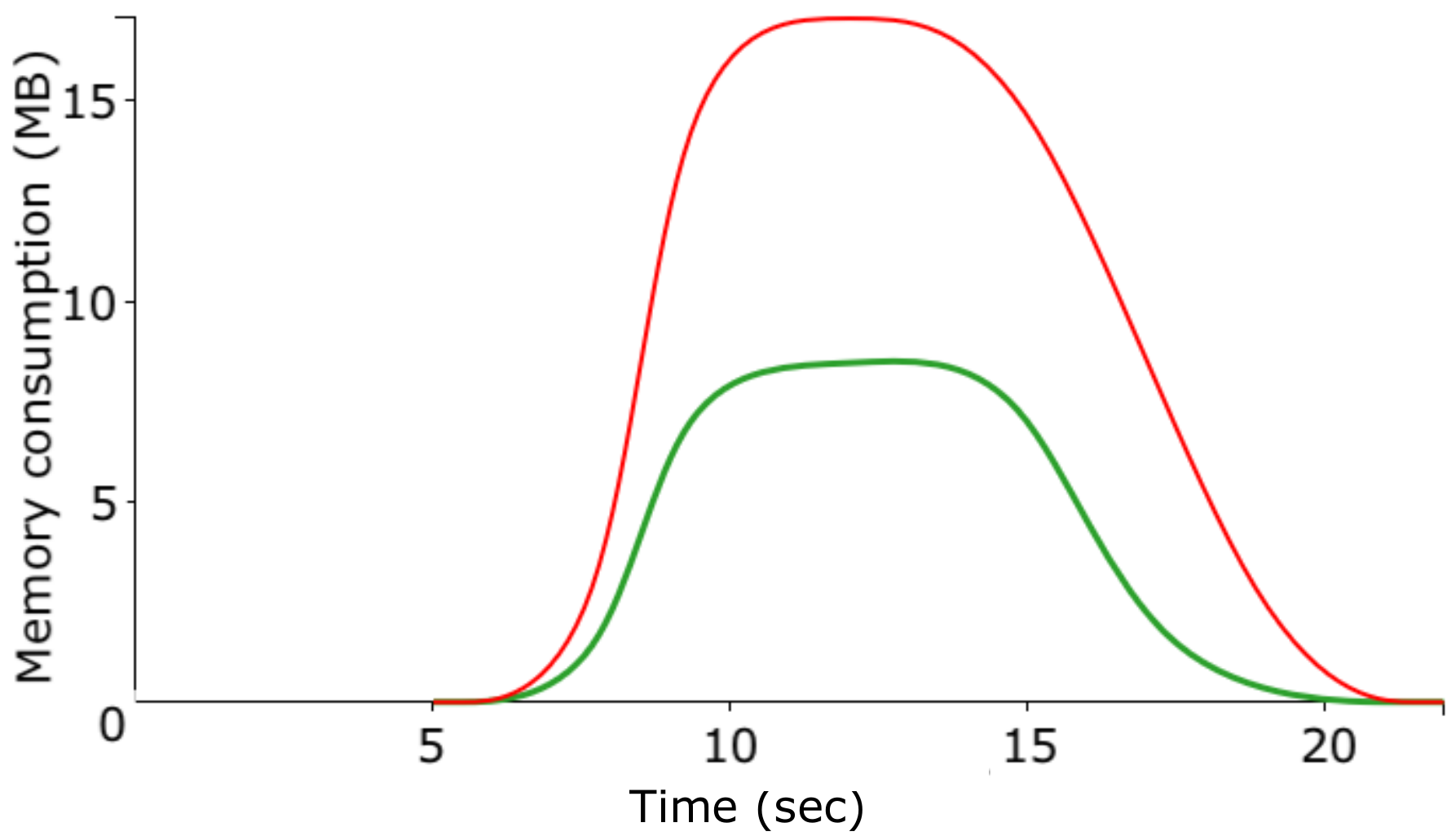}
    \caption{Memory consumption  of our malware PoC (Argon2d) Malware:red line, System Idle (PID 0):green line.}
    \label{fig:mem_footprint}
    \end{minipage}
    \hfill
    \begin{minipage}[t]{0.32\textwidth}
    \centering
    \includegraphics[width=1\columnwidth]{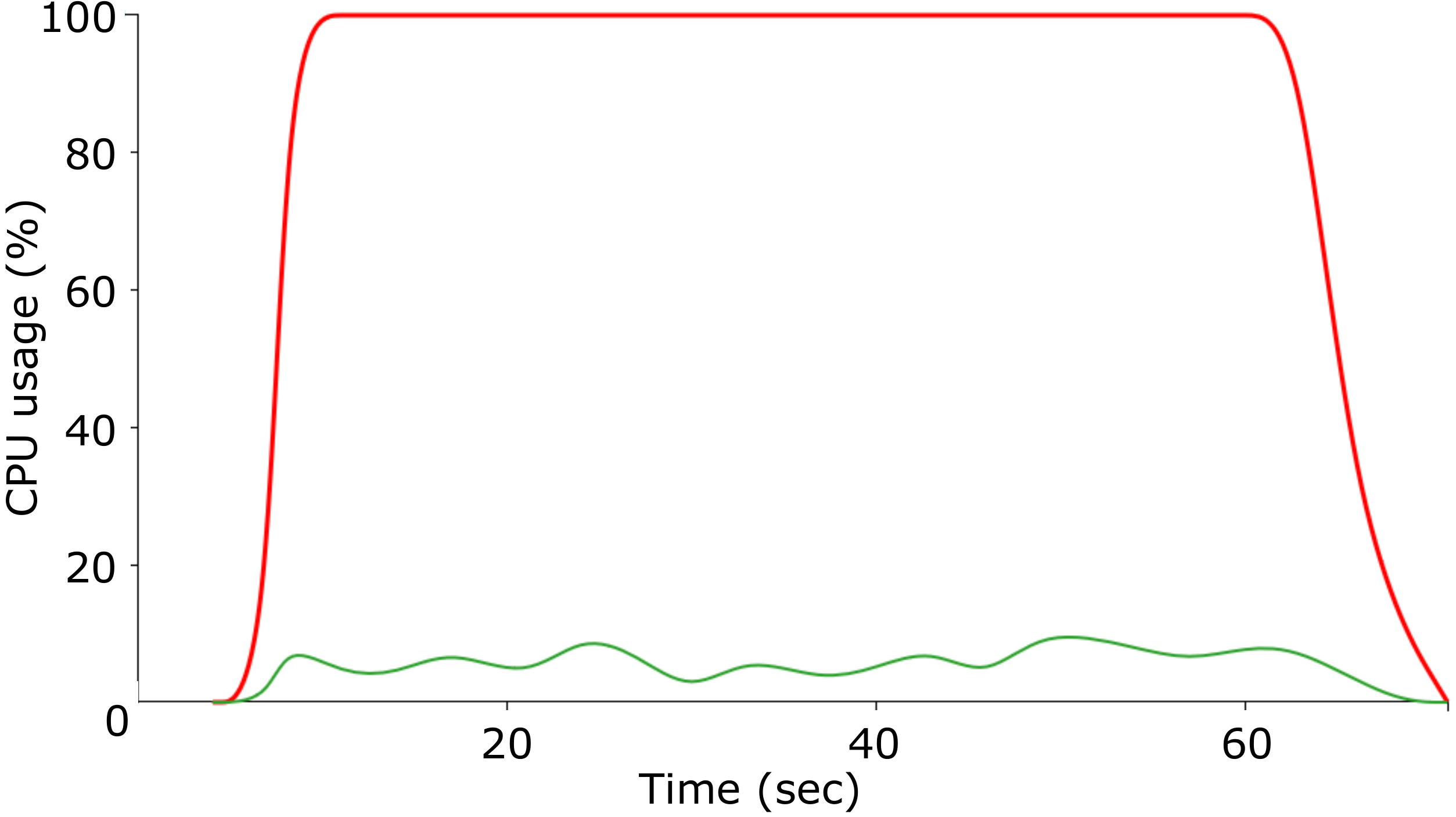}
    \caption{CPU consumption of our malware PoC. T=60 seconds and 0.5 seconds between each PoW execution. Malware:red line, System Idle (PID 0):green line.}
    \label{fig:evader_ant2}
    \end{minipage}\vspace{-0.4cm}
\end{figure}




\point{CPU and memory usage}
The main downside of associating a PoW with a malware sample is an increase in both CPU and memory consumption.  We here report on CPU and memory consumption as measured by our sandbox. 
Figures~\ref{fig:cpueva} and~\ref{fig:mem_footprint}  compare, respectively, CPU and memory utilization of our malware (red line) with System Idle (PID 0). With respect to  CPU usage, the PoW associated with our malware causes an (expected) 100\% utilization for the whole duration of the PoW ($T=10$~sec). With respect to  memory utilization, our malware only requires about 17~MB versus the 7~MB that utilizes a sample system process like System Idle (PID 0). This is a minor increase, unlikely to raise any suspicion. 


Next, we investigate whether we can reduce the CPU usage of our PoC ransomware
by setting a longer $T$ (\eg 60~sec) and a sleep of $0.5$~sec between each PoW execution. Despite such sleeps, Figure~\ref{fig:evader_ant2} still shows 100\%  CPU utilization for the whole $T$ (60~sec in this test). The lack of CPU reduction associated with the extra sleeps is counter-intuitive. The likely explanation is that the  sandbox leverages a coarse CPU monitoring tool and, thus, the CPU reduction associated with our extra sleeps gets averaged out. 
These results provide a foundation to detect evasion techniques based on PoW. A sandbox could 
attempt heuristics based on a binary's CPU and memory consumption. We argue, however, that this is quite challenging because of the potential high number of false positives that can be generated.

\section{Countermeasures}
\label{sec:defences}

Evasion techniques are easily comparable with other anti-analysis techniques like \emph{packing}. Packing techniques have evolved to such sophistication that it has become practically impossible to unpack a malware sample without dynamically executing it~\cite{lineage,ugarte}. However, dynamically executing a sample can indeed trigger evasion techniques like stalling code. To counter evasion techniques, and especially the ones that \name implements, one idea would be to fingerprint the algorithms, \eg CPU and memory footprint. However, it would be very easy for attackers to apply code polimorphism techniques and produce variants that diverge from the original implementation, as it is done with packers. This will constitute a challenge for the sandbox, which could generate a false negative by not being able to spot the algorithm. In Table~\ref{tab:caseStudies},  the Hello World program is detected as malicious and our technique reduces its detection rate and with a code virtualizer it makes the sample completely stealth. 

\point{Fingerprinting evasion}
A common solution against red pills~\cite{fistful} is to reduce the amount of instructions failing due to emulation. As Martignoni et al.~\cite{emufuzzer,kemufuzzer} show, the analysis can be automated and the fixes can be easily produced. However, with PoW the computational model is not seeking for emulation/virtualization failures or malfunctions. Instead, PoW is acting as a probe to spot a side channel in the execution time of the algorithm, which in this case is time-based.

\point{Virtualized instructions set} Native execution of the cryptographic instructions is another potential countermeasure that could be considered to mitigate our approach.
In such a case, the cryptographic instructions of the PoW algorithm are not emulated by the sandbox environment, but directly executed on the native CPU. Avoiding the emulation of the cryptographic instructions could clearly improve the computational performance of the PoW algorithm and reduce the success probability of the evasive behavior showed by \name.
The technique described in the Inspector Gadget paper~\cite{DBLP:conf/sp/KolbitschHKK10}, which works at the program analysis level, may also work to avoid the execution of our evasion code. Once the sample is unpacked, it would be possible to extract and execute only the malware branch of the code as a gadget and analyze its behavior in isolation. However, a sufficiently complex packer or emulator would make such process very tedious and require manual effort, which makes this solution excessively complex to be implemented in an automated malware analysis service.

\point{Specialized hardware}
Even if our choice, Argon2, is resilient to specialized circuits for mining (ASICs and FPGAs), other PoW algorithms are not, and hence an analyst could equip his sandbox with a miner~\cite{antminer}. Such a dedicated hardware is expensive for a non-professional user (around $\$3,000$ at the time of writing). Nonetheless, if the phenomenon of sandbox evasion due to PoW proliferate, having such a platform would be of great help to offload the PoW calculations, through a tailored interface, and continue the execution of the malware sample inside the sandbox. The cost/benefit trade-off of adopting such a measure really depends on the intended scale of the analysis platform. For example, according to VirusTotal statistics~\cite{virustotalstats}, the service receives weekly more than 3M PE binaries. Hence, a dedicated hardware to defeat PoW evasion based techniques  seem a good compromise, since it allows to analyze and discover new malicious behaviors.

\point{Spoofing timers}
The sandbox that gets a \name-ed malware could try to delay the time, which could mean to make our $T=10$ seconds last much longer to achieve the payload execution. This approach may work well. Though, if we expect a total of at least 50 PoW iterations (see Section~\ref{sec:framework:estim}) and the sandbox is not able to execute more than one in about a minute for a unique malware sample, the analysis would take more than one hour. This will eventually extract the payload that will then require extra work to be reverse engineered, understood, and fingerprinted. Hence, this approach may not scale in terms of time/cost for the large number of samples that online sandboxes analyze daily.

\point{Bare-Metal Sandboxes} Using bare metal hardware represents a reasonable solution that might be adopted within corporate companies but it is not possible to use such technology at Internet scale, \ie cloud-based solutions like Virus Total. Also, isolated sandboxes do not benefit of the information that on-line in cloud services have which leverages large scale cross-correlations.
\section{Discussion}

\subsection{Ethical Considerations}
\label{sec:ethics}
The results obtained by \name regarding the analyzed publicly available sandboxes, normally used by malware analysts under their term of service (ToS), demonstrate that our technique works consistently either in our custom Cuckoo Sandbox implementation or in proprietary solutions. Our aim, though, is not to disrupt any business nor to difficult the operation of companies that profit from providing malware behavior analysis. We contacted all the platforms and vendors that we have tested with \name and we notified them about our findings. Part of the vendors were very positive and agreed to further collaborate to work on practical countermeasures. 
Unfortunately, the response we received from other vendors opposed any dissemination of our results, adopting a shortsighted security-through-obscurity approach which is not novel in our community.
Consequently, tested vendors have been anonymized to avoid violation of their ToS. We purposely maintained the number of new variants submitted to the bare minimum, but our approach may transform easily any existing sample into a new one. The authors are available for contact for further information disclosure.

\subsection{Bare-Metal Environments}
In~\cite{barecloud} the authors present BareCloud a bare-metal system which helps to detect evasive malware. This system in order to execute malware trades visibility against transparency. In other words it makes the analysis system transparent (non-detectable by malware) and produces less powerful analysis data (limited instrumentation). Indeed their detection technique leverages hierarchical similarity~\cite{kdd} comparison between different malware execution traces (virtualized and emulated) systems i.e., (Ether~\cite{ether}, Anubis~\cite{anubis}, and VirtualBox~\cite{cuckoo}). One of the biggest problem of hierarchical similarity algorithms is scalability, which means that the algorithm should be polynomial in time and space. An example~\cite{simil} of application and analysis of hierarchical similarity for binary program comparison shows $O(n^2)$ complexity. Hence using BareCloud as a production system for example for VirusTotal which claims~\cite{virustotalstats} about 1.5M daily submissions means that the hierarchical comparison would approximate 2.250 billion of operations daily to detect evasive malware with bare metal equipment. It is evident that BareCloud can be useful in special cases, as briefly stated above, where also a manual analyst can make the difference. For the sake of scalability though virtualization and emulation methods cannot be fully replaced, even if it would be possible to instrument in hardware an entire system~\cite{10.1145/2818000.2818030}, the approach would suffer many other issues, for instance having a lot of physical hardware and maintaining it.
%
%
%
\subsection{Economical denial of sustainability} Online sandboxes, like any other business, have costs to sustain. Ignoring evasive malware to avoid an additional cost is (for now) understandable. Unfortunately, malware that exploits \name's technique implies additional energy and memory costs, especially if submitted in large scale to such systems, opening avenues to EDoS attacks, which will try to make the on-line service not sustainable economically. These on-line services receive on average 1.5M samples daily. It is not difficult to imagine how much energy just a tenth of the total submissions can consume if it is running PoW. Such algorithm is one of the most energy intensive operation that a computer can perform. For instance, the yearly energy consumption of Bitcoin's blockchain is comparable to the one of a country such as Tunisia or Czech Republic~\cite{bitpower}. We strongly recommend that not all evasion techniques are the same, and every technique that exploits hardware consumption side channels should be properly analyzed to avoid service disruption.
\section{Related Work}
\label{sec:related}
There is a significant body of research~\cite{c1,c2,c3,canali,DBLP:conf/ccs/LanziBKCK10,graziano,rotalume} focusing on both designing novel evasion techniques for malware and also providing mechanisms to detect them. We next discuss the most relevant works related to ours.
%
%
\point{Fingerprinting emulated environments}
By recognizing the sandboxes of different vendors, malware can identify
the distinguishing characteristics of a given emulated environment and alter its behavior accordingly. 
The work in~\cite{1624022} introduced the notion of \emph{red  pill} and released a short exploit code snippet that could be used to detect whether the code is executed under a VM or in a real platform.  In~\cite{fistful}, the authors propose an automatic and systematic technique (based on EmuFuzzer~\cite{emufuzzer}) to generate red pills for detecting whether a program is executed inside a CPU emulator.
%
In~\cite{kemufuzzer}, the authors build KEmuFuzzer, which leverages protocol-specific fuzzing and differential analysis. KEmuFuzzer forces the hosting virtual machine and the underlying physical machine to execute specially crafted snippets of user- and system-mode code before comparing their behaviors. 
In~\cite{blackthorne} authors presented AVLeak, a tool that can fingerprint emulators running inside commercial antivirus (AV) software, which are used whenever AVs detect an unknown executable. The authors developed an approach that allows them to deal with these emulators as black boxes and then use side channels for extracting fingerprints from each AV engine.
Instead, we show that even with completely transparent analysis programs, the real environment can be used by the malware to determine that it is under analysis.
In~\cite{spotless_sand} authors propose a ML-based approach to detect emulated environments. This technique is based on the use of features such as the number of running processes, shared DLLs, size of temporary files, browser cookies, etc. These features are named by the authors ``wear-and-tear artifacts'' and are present in real system as opposed to sandboxes. The authors use such features to train an SVM classifier. We also rely on modeling a distinguishing feature, in our case is a time channel arising from the asymptotic behavior of a Pow, not the presence or absence of system artefacts.



In~\cite{franklin2008remote}, authors introduce the virtual machine monitor (VMM) detection and they propose a fuzzy benchmark approach that works by making timing measurements of the execution time of particular code sequences executed on the remote system. The fuzziness comes from heuristics which they employ to learn characteristics of the remote system's hardware and its configuration. 
In~\cite{chen2008towards}, the authors present a technique that leverages TCP timestamps to detect anomalous clock skews in VMs. A downside of the approach is that it requires the transmission of streams of hundreds of SYN packets to the VM, something that can be detected in the case of a honeypot VM and flagged as malicious behavior. 
Compared to the previous approaches, \name is more principled and offers a solid basis founded on cryptographic primitives (PoW) with a predictable and reproducible computational behavior on different tested platforms.

\point{Detecting evasive malware}
In~\cite{dinaburg2008ether}, the authors propose Ether, a malware analyzer that eliminates in-guest software components vulnerable to detection. Ether leverages hardware virtualization extensions such as Intel VT, thus residing outside of the target OS environment. In \cite{barecloud}, the authors present an automated evasive malware detection system based on bare-metal dynamic malware analysis. Their approach is designed to be transparent and thus robust against sophisticated evasion techniques. The evaluation results showed that it could automatically detect 5,835 evasive malware out of 110,005 tested samples. In~\cite{balzarotti2010efficient}, authors propose a technique to detect malware that deploys evasion mechanisms. Their approach works by comparing the system call trace recorded when running a malware program on a reference system with the behavior observed in the analysis environment. In~\cite{lindorfer2011detecting}, authors propose a system for detecting environment-sensitive malware by comparing its behavior in multiple analysis sandboxes in an automated way. Compared to previous techniques, our approach is agnostic to system artifacts and cannot be recognized by only monitoring the system operations.

\section{Conclusion}
\label{sec:conclusion}
Online malware scanning services are becoming more and more popular, allowing users to upload and scan artefacts against AV engines and malware analysis sandboxes.
Common mechanisms used by malware samples to avoid detection include the inspection of signals that imply the existence of a virtualized or emulated environment. These strategies triggered an arms-race where online malware scanners patch such signals to make virtualization transparent.
In this paper, we leverage PoW techniques as the basis for a novel malware evasion technique due to their ability to fingerprint real hardware. We provide empirical evidence of how it can be used to evade online malware analysis sandboxes and discuss potential countermeasures. The implementation of our approach goes beyond a simple proof-of-concept, showing that injecting evasion modules can be easily automated on any arbitrary sample. We make our code and results publicly available in an attempt to increase reproducibility and stimulate further research in this area.

\bibliographystyle{Preamble/splncs04}
\bibliography{main}
\end{document}